\documentclass[preprint]{aastex}

\usepackage{amsmath}

\newcommand{\Teff}{$T_{\rm eff}$}
\newcommand{\logg}{log\,\textit{g}}
\newcommand{\vsini}{\textit{v}\,sin\,\textit{i}}

\DeclareRobustCommand\ion[2]{%
  \mbox{#1\kern0.2em%
  \smaller\rmfamily%
  \edef\@tempa{\@car#2\@nil}%
  \ifcat1\@tempa%
  \@Roman{#2}%
  \else%
  \uppercase{#2}%
  \fi}}

\def\ion#1#2{\ensuremath{\mathrm{#1}\;}{\protect\small\rm{#2}}}
\DeclareRobustCommand\nodata{$\cdots$}

\def\sn1{${\rm s}^{-1}$}

\shorttitle{Algorithm to fit synthetic stellar spectra}
\shortauthors{Fierro-Santill\'an et al.}

\begin{document}

\bibliographystyle{apjsty}

\title{FIT{\it spec}: a new algorithm for the automated fit of synthetic stellar spectra for OB stars}

\author{Celia R. Fierro-Santill\'an$^{1}$, Janos Zsarg\'o$^{2}$, Jaime Klapp$^{1,3}$,
Santiago A. D\'{\i}az-Azuara$^{4}$, Anabel Arrieta$^{5}$, Lorena Arias$^{5}$, 
Leonardo Di G. Sigalotti$^{6}$}
\affil{$^1${\sc abacus}-Centro de Matem\'atica Aplicada y C\'omputo de Alto Rendimiento,
Departamento de Matem\'aticas, Centro de Investigaci\'on y de Estudios Avanzados
(Cinvestav-IPN), Carretera M\'exico-Toluca km. 38.5, La Marquesa, 52740 Ocoyoacac, Estado
de M\'exico, Mexico}
\affil{$^2$Escuela Superior de F\'{i}sica y Matem\'aticas, Instituto Polit\'ecnico Nacional
(IPN), Luis Enrique Erro S/N, San Pedro Zacatenco, 07738 Gustavo A. Madero, Ciudad de M\'exico,
Mexico}
\affil{$^3$Departamento de F\'{\i}sica, Instituto Nacional de Investigaciones Nucleares (ININ),
Carretera M\'exico-Toluca km. 36.5, La Marquesa, 52750 Ocoyoacac, Estado de M\'exico, Mexico}
\affil{$^4$Instituto de Astronom\'{i}a, Universidad Nacional Aut\'onoma de M\'exico (UNAM),
Avenida Universidad, 04510 Ciudad de M\'exico, Mexico}
\affil{$^5$Universidad Iberoamericana, Prolongaci\'on Paseo de la Reforma 880, Santa Fe,
Contadero, 01219 Ciudad de M\'exico, Mexico}
\affil{$^6$\'Area de F\'{\i}sica de Procesos Irreversibles, Departamento de Ciencias B\'asicas,
Universidad Aut\'onoma Metropolitana-Azcapotzalco (UAM-A), Av. San Pablo 180, 02200
Mexico City, Mexico}

\begin{abstract}
In this paper we describe the FIT\textit{spec} code, a data mining tool for the automatic 
fitting of synthetic stellar spectra. The program uses a database of 27\,000 {\sc cmfgen} 
models of stellar atmospheres arranged in a six-dimensional (6D) space, where each dimension 
corresponds to one model parameter. From these models a library of 2\,835\,000 synthetic 
spectra were generated covering the ultraviolet, optical, and infrared region of the 
electromagnetic spectrum. Using FIT\textit{spec} we adjust the effective temperature and the
surface gravity. From the 6D array we also get the luminosity, the metallicity, and three 
parameters for the stellar wind: the terminal velocity ($v_\infty$), the $\beta$ exponent 
of the velocity law, and the clumping filling factor ($F_{\rm cl}$). Finally, the projected 
rotational velocity ($v\cdot\sin i$) can be obtained from the library of stellar spectra. 
Validation of the algorithm was performed by analyzing the spectra of a sample of eight O-type 
stars taken from the {\sc iacob} spectroscopic survey of Northern Galactic OB stars. The 
spectral lines used for the adjustment of the analyzed stars are reproduced with good accuracy. 
In particular, the effective temperatures calculated with the FIT\textit{spec} are in good 
agreement with those derived from spectral type and other calibrations for the same stars. 
The stellar luminosities and projected rotational velocities are also in good agreement 
with previous quantitative spectroscopic analyses in the literature. An important advantage 
of FIT\textit{spec} over traditional codes is that the time required for spectral analyses 
is reduced from months to a few hours.
\end{abstract}

\keywords{Astronomical databases: miscellaneous -- Methods: data analysis -- 
Stars: atmospheres -- Stars: massive -- Radiative transfer} 

\section{Introduction} 

The self-consistent analysis of spectral regions from the ultraviolet (UV) to the 
infrared (IR) radiation band has been made possible because of the large amount of 
publicly available data combined with the existence of sophisticated stellar atmosphere 
codes such as {\sc cmfgen} \citep{Hillier98}, {\sc tlusty} \citep{Hubeny95}, and 
{\sc fastwind} \citep{Santolaya97,Puls05}. As a result of this, significant advances have 
been made toward understanding the physical conditions prevailing in the atmospheres and 
winds of massive stars. For instance, \cite{Fullerton00} showed that there were 
inconsistencies in the optical effective temperature scale in the early far-UV spectra 
when compared with the scale implied by the observed wind ionization. On the other hand, 
studies conducted by \cite{Martins02} and \cite{Martins03} have shown that the neglect of 
line blanketing in the models leads to a systematic overestimate of the effective
temperature when derived from optical H and He lines. An improvement over these 
previous calibrations was reported by \cite{Martins05}, where a detailed treatment of 
non-LTE line-blanketing in the expanding atmospheres of massive stars was taken into 
account. After direct comparison to earlier calibrations of \cite{Vacca96}, they found
effective temperature scales of dwarfs, giants, and supergiants that were lower from 
2000 to 8000 K, with the reduction of temperature being the largest for the earliest 
spectral types and for supergiants. The luminosities were also reduced by 0.20 to 0.35 
dex for dwarfs, about 0.25 dex for all giants, and by 0.25 to 0.35 dex for supergiants, 
with these reductions being almost independent of spectral type for the latter two cases. 
A more recent analysis by \cite{Martins15}, using the {\sc cmfgen} code with line-blanketing
included, found effective temperatures of Galactic O stars that were in good agreement
with the {\sc fastwind} values reported by \cite{Simon14} and \cite{Simon17}.
On the other hand, \cite{Crowther02}, \cite{Hillier03}, and \cite{Bouret03} have 
simultaneously performed analyses of $FUSE$, $HST$, and optical spectra of O-type stars 
and were able to derive consistent effective temperatures using a wide variety of 
diagnostics.

A further important result was the recognition of the effects of wind inhomogeneities 
(i.e., clumping) on the spectral analyses of O-type stars. For instance, \cite{Crowther02} 
and \cite{Hillier03} were unable to reproduce the observed \ion{P}{V} $\lambda\lambda$1118-1128 
profiles when using mass-loss rates derived from the analysis of H$\alpha$ lines. The 
only way the \ion{P}{V} and H$\alpha$ profile discrepancies could be resolved was by 
either assuming substantial clumping or using unrealistically low phosphorus abundances. 
Therefore, as a consequence of clumping, the mass-loss rates have been lowered by factors 
ranging from $\sim 3$ to 10. Moreover, new observational clues to understand 
macroturbulent broadening in massive O- and B-type stars have been provided by
\cite{Simon17}. They found that the whole O-type and B supergiant domain is dominated
by massive stars ($M_{\rm ZAMS}\gtrsim 15$ M$_{\odot}$) with a remarkable non-rotational
line broadening component, which has been suggested to be a spectroscopic signature of
the presence of stellar oscillations in those stars.

Conduction of the above investigations with the aid of any of the existing stellar atmosphere 
codes is by no means a simple task. Running these codes and performing reliable analyses 
and calibrations demand a lot of experience that unfortunately many researchers may
have no time to acquire. On the other hand, for each interpolation, several models
need to be ran, which takes a long time. Thus, the program for fitting
atmospheric parameters spends most of its time with this. Therefore, it is desirable
to optimize the calculation by developing databases of pre-calculated models as well
as the tools that are necessary for their use. Such databases will allow astronomers
to save time and analyze stellar atmospheres with reasonable accuracy and without the 
need of running time consuming simulations. Furthermore, these databases will also 
speed up the study of a large number of observed spectra that are still waiting for 
analysis.

The basic parameters of such databases of pre-calculated models are: the surface 
temperature (\Teff), the stellar mass ($M$), and the surface chemical composition. 
However, an adequate analysis of massive stars must also take into account the
parameters associated with the stellar wind, such as the terminal velocity ($v_\infty$), 
or the mass-loss rate ($\dot M$), and the line clumping. If we take into account the 
variations of all necessary parameters, the number of pre-calculated models that are 
actually needed will increase exponentially. Therefore, production of such databases 
is only possible through the use of supercomputers.

At present, there are a few databases of synthetic stellar spectra available and only 
with a few tens or hundreds of stellar models (see, for example, \cite{Fierro15}, or the 
{\sc pollux} database \citep{Palacios10}). To improve on this we have developed 
a database with tens of thousands of stellar models \citep{Zsargo17}, which we will
release for public use in a short time. Since it is impossible to manually compare an
observed spectrum with such an amount of models, it is imperative to develop appropriate
tools that allow the automation of this process without compromising the quality of the 
fitting. With this in mind, we have created FIT\textit{spec}, which is a program that 
searches in our database for the model that better fits the observed spectrum 
in the optical. This program uses the Balmer lines to measure the surface gravity 
($\log g$) and the line ratios \ion{He}{II}/\ion{He}{I} to estimate \Teff. The
paper is organized as follows. In \S 2, we describe the grid and the 
six-dimensional (6D) parameter space of the model database. In \S 3, we give a detailed 
description of the algorithm and in  \S 4, we test the algorithm by analyzing the 
spectra of a sample of eight O-type stars taken from the {\sc iacob} spectroscopic 
database of Northern Galactic OB stars. Finally, in \S 5 we summarize the main conclusions.

\section{Model database in a six-dimensional parameter space}

The stellar models are calculated using the more sophisticated and widely used non-LTE 
stellar atmosphere code {\sc cmfgen} \citep{Hillier98}. The code calculates the full spectrum 
and has been used successfully to model OB stars, W-R stars, luminous blue variables, 
and even supernovae. It determines the temperature, the ionization structure, and the 
level populations for all elements in the stellar atmosphere and wind. It solves the 
radiative transfer equations in the co-moving frame in conjunction with the statistical 
and radiative equilibrium equations under the assumption of spherical symmetry. The 
hydrostatic structure can be computed below the sonic point, thereby allowing for the 
simultaneous treatment of spectral lines formed in the atmosphere, the stellar wind, and 
the transition region between the two. In particular, the code is well suited for the 
study of massive OB stars with winds.

At present, our database contains 27\,000 atmosphere models, arranged in a 6D space. When 
all parameter combinations are taken into account, we then expect to have 80\,000 
models. Each dimension in 6D space 
corresponds to one parameter of the model. In addition to the surface temperature (\Teff), 
the luminosity ($L$), and the metallicity ($Z$) of the star, we consider three more 
parameters for the stellar wind, namely the terminal velocity ($v_\infty$), the $\beta$ 
exponent of the velocity law, and the clumping filling factor ($F_{\rm cl}$). Here by 
$v_\infty$ we mean the velocity of the stellar wind at a large distance from the star. 
Outside the photosphere, we model the wind velocity as a function of the stellar radius 
using the $\beta$-type law \citep{Cassinelli79}, i.e.
\begin{equation}
v(r)=v_{\infty}\left(1-\frac{r}{R_{\star}}\right)^\beta,
\end{equation}
where the free parameter $\beta$ controls how the stellar wind is accelerated to reach 
the terminal velocity. Low values of $\beta$ (i.e., $\beta=0.8$) indicate a fast wind 
acceleration, while high values (i.e., $\beta=2.3$) indicate lower accelerations. Since
the stellar wind is not necessarily homogeneous, we assume that it contains gas in the
form of small clumps or condensations. The volume filling factor $F_{\rm cl}$ is then 
the fraction of the total volumen occupied by the gas clumps, while the space between 
them is assumed to be a vacuum. In addition, the mass-loss rate that is used for
the models is taken from the evolutionary tracks of \cite{Ekstrom12}.

In Table 1 we list the values of the relevant model parameters. However, not all of
them are truly free parameters. For instance, some of them are associated to other parameters
(i.e., the mass loss rate, $\dot{M}$, is completely determined when \Teff, \logg, and $Z$
are known, while $R$ and \logg\hspace{0.01cm} depend on $M$ and $L$). The dependence of
other parameters such as $\beta$ and $F_{\rm cl}$ has not been sufficiently explored and
so they could be degenerate with other parameters. For each model, we calculate the 
synthetic spectra in the UV (900-3\,500 \AA), optical (3\,500-7\,500 \AA), and near IR 
(7\,500-30\,000 \AA) radiation bands. In order to facilitate comparison with the observations, 
the synthetic spectra are rotationally broadened using the program {\sc rotin3} \citep{Hubeny95}, 
with rotational velocities between 10 and 350 km\,s$^{-1}$ separated by intervals of 10 
km\,s$^{-1}$. These discrete values result in a library of  27\,000 models $\times$ 
3 bands $\times$ 35 values of the rotational velocity $=$ 2\,835\,000 synthetic spectra. 
The main parameters of any model atmosphere are the luminosity ($L$) and the effective 
temperature (\Teff) from which we can determine the location of the star in the H-R diagram. 
As appropriate constraints to the input parameters, we use the evolutionary tracks of 
\cite{Ekstrom12} calculated with solar metallicity ($Z=0.014$) at the zero age of the main 
sequence (ZAMS). Each point of a track corresponds to a star with specific values of 
\Teff, luminosity ($L$), and stellar mass (\textit{M}). We have calculated several
models along each track at approximate discrete intervals of 2\,500\,K in \Teff. With
this provision, the stellar radius, $R$, and the surface gravity, \logg,\ were 
calculated to determine the luminosity, \textit{L}, and the stellar mass, \textit{M}, 
corresponding to the track. The terminal velocities of the O-type stars in our sample
are fitted by {\bf$v_{\infty}=2.1v_{\rm esc}$}, where $v_{\rm esc}$ is the photospheric
escape velocity. The chemical elements that are taken into account in our models
are H, He, C, N, O, Si, P, S, and Fe. In particular, the values of the first five 
elements are taken from \cite{Ekstrom12}, while for consistency we take the solar 
metallicity reported by \cite{Asplund09} for Si, P, S, and Fe in all models.

The code {\sc cmfgen} employs the concept of ``super levels'' for the atomic models, where 
levels of similar energy are grouped together and treated as a single level in the 
statistical equilibrium equations \citep[see][and references therein]{Hillier98}. The 
stellar models in this project include 28 explicit ions of the different elements as 
a function of their \Teff. Table 2 summarizes the levels and super levels that are included 
in the models. The atomic data references are given in the appendix of \cite{Herald04}.

\section{The FIT\textit{spec} Algorithm}

An experienced astronomer can make a qualitative fit by comparing by eye one or more 
models with the observed spectrum. However, this procedure becomes too cumbersome and time
consuming if hundreds of models must be compared. Also, when the number of models is too large, 
the objectivity can be easily compromised. FIT\textit{spec} is a heuristic tool that mimics 
the procedure followed by an experienced
astronomer to analyze observed stellar spectra. Due to the big size of the database, 
it is basically impossible to manually compare the observed spectra with the available 
models and find the best fit. For this purpose we have developed FIT\textit{spec}, which
was also designed to perform this task in a much shorter time compared to traditional 
fitting algorithms. The usual method to estimate \Teff\ 
in stellar atmospheres is to compare the equivalent widths (EW) of two lines of the same 
element in consecutive states of ionization. We have adopted four EW ratios of \ion{He}{II} 
and \ion{He}{I} lines to estimate \Teff, namely
\begin{equation}
\frac{EW(He II \, \lambda 4541)}{EW(He I \, \lambda 4471)} \; (a), \; \; \;
\frac{EW(He II \, \lambda 4200)}{EW(He I \, \lambda 4026)} \; (b), \; \; \;
\frac{EW(He II \, \lambda 4200)}{EW(He I \, \lambda 4144)} \; (c), \; \; \;
\frac{EW(He II \, \lambda 4541)}{EW(He I \, \lambda 4387)} \; (d). \label{eqn:ratios} 
\end{equation}
For comparison, \cite{Walborn90} used the ratios \ref{eqn:ratios}(a) and (b) 
to classify O3-B0 main sequence stars, while the ratios \ref{eqn:ratios}(c) and 
(d) were suitable only for the classification of the later types in 
the same range. In contrast, we have recorded these ratios for all models in our 
database. To use FIT\textit{spec}, the user must provide the observed EW of
\ion{He}{II} $\lambda\lambda$ 4541, 4200; \ion{He}{I}
$\lambda\lambda$ 4471, 4387, 4144; and \ion{He}{I}+\ion{He}{II} $\lambda\lambda$ 4026 as 
input data. These values can be easily measured by any astronomical software as, for 
example, {\sc iraf}\footnote{{\sc iraf} is written and supported by 
the National Optical Astronomy Observatories (NOAO) in Tucson, Arizona. NOAO is operated 
by the Association of Universities for Research in Astronomy (AURA), Inc. under 
cooperative agreement with the National Science	Foundation (NSF).}. The algorithm then 
calculates the EW ratios for the observed lines and compares them with those for the 
models in the database.

Under the assumption that the best fit model is the one that accurately reproduces the 
ratios for the observed spectrum, we can calculate the differences between the observed 
ratios and those pertaining to each model in terms of the relative error
\begin{equation}
Error\left(\frac{\ion{He}{II}}{\ion{He}{I}}\right) =
\frac{\left(\frac{\ion{He}{II}}{\ion{He}{I}}\right)_{\rm obs} -
\left(\frac{\ion{He}{II}}{\ion{He}{I}}\right)_{\rm mod}}
{\left(\frac{\ion{He}{II}}{\ion{He}{I}}\right)_{\rm obs}},
\end{equation}
where $\left(\ion{He}{II}/\ion{He}{I}\right)_{\rm obs}$ is any of the ratios
calculated from the observed spectrum and
$\left(\ion{He}{II}/\ion{He}{I}\right)_{\rm mod}$ is the corresponding ratio
for a model. The metric defined by Eq. (3) provides a good measure of the
difference between model and observation.

FIT\textit{spec} then calculates a weighted average of the errors in the four ratios 
considered. The weight of each ratio is an input parameter and must be provided by 
the user. The program was designed to find all models with average errors less than 
50\%, save the description of these models in an output file, and produce a 
graphical output to visualize the location of the models in the 6D parameter space 
(see Fig. 1). The next step consists of estimating the surface gravity, $\log g$. For 
this purpose the EWs of the \ion{H}{I} Balmer lines are used. The errors in the EWs of 
the six Balmer lines ($\lambda\lambda $3835, 3889, 3970, 4102, 4341, 4861) are then 
calculated, which are finally used to estimate the $\log g$ of the star. As for the 
spectral type - effective temperature (SpT - \Teff) calibration, these errors are 
calculated using the metric
\begin{equation}
Error_{\rm EW}=\frac{EW_{\rm obs} - EW_{\rm mod}}{EW_{\rm obs}}
\end{equation}
where $Error_{\rm EW}$ indicates how different a model and the observation are for a 
specific line. FIT\textit{spec} also calculates the weighted averages of the relative errors 
in the EWs of the Balmer lines where the weights must be provided by the user. The 
algorithm first finds out the models that are within an error of 50\% and then picks
those models that have both $Error(\ion{He}{II}/\ion{He}{I})$ and $Error_{\rm EW}$ less
than 50\%. The total error is calculated according to
\begin{equation}
Error_{\rm tot}=\sqrt{\left(Error_{\rm EW}\right)^{2} + 
	\left[Error\left(\frac{\ion{He}{II}}{\ion{He}{I}}\right)\right]^{2}}.
\end{equation}
Finally, the program sorts the errors from lowest to highest values of $Error_{\rm tot}$
and generates a file containing only those models whose total error is less than 10\% (see Fig. 
2). The sequence of steps followed by FIT\textit{spec} is shown in the flowchart of Fig. 3.

\section{Results and discussion}

Validation of FIT\textit{spec} is made by testing the algorithm for a sample of eight 
O-type main-sequence stars. These stars were chosen on the basis of having known
available observational data and spectral classification in the {\sc iacob} database
\citep{Simon11}. In addition, they are located in a region of the H-R diagram where 
our database has the highest density of models. Table 3 lists the stellar parameters
as obtained from the best fitted models as found by FIT\textit{spec} for each selected
star in our sample. Uncertainties in the effective temperature and luminosity
are 1 kK and 0.15 dex, respectively, for all models in the sample. This uncertainties 
were estimated from the models themeselves. The errors in \Teff\ take into account the 
models that fit reasonably well the EWs of the He I and He II lines in a global way, while 
the errors in $L$ include the models that fit reasonably well the EWs of the Balmer lines in 
a global way. In addition, the first and second columns of Table 4 list the selected stars 
and their spectral type, respectively, while the next eight columns compare their effective 
temperatures, surface gravities, and luminosities (also listed in Table 3) with the 
corresponding values from spectral type calibrations and other spectral analyses reported in 
the literature. Finally, the last two columns in Table 4 compare the projected rotational 
velocities as found by FIT\textit{spec} with the corresponding values obtained from several 
other spectral calibrations, as indicated by the references listed in the footnote of Table 4.

In passing, we note that star HD54662 is a peculiar object. Some authors have treated 
it as if it were a single star \citep{Markova04,Krticka10}, although there is enough evidence 
that it is actually a binary star \citep{Fullerton90,Sana14,Mossoux18}. However, in this work 
HD54662 was taken as a single star because the spectrum extracted from the {\sc iacob} 
database shows no evidence of binarity. The corner plots of Figure 1 show the distribution 
in the 6D space of the models with relative errors less than 50\% in the  He II/He I ratios 
for star HD54662 of spectral type O7 V, while Fig. 2 shows the distribution of the models when
$Error_{\rm tot}\lesssim 10$\% for the same star. When only the temperature criterion is 
considered, the models span a finite range of values of the parameter space. That is, for
prescribed values of the metallicity, filling factor, and $\beta$ exponent of the velocity
law, the effective temperature and luminosity of the star can have different values over
a finite range. However, when the more stringent criterion $Error_{\rm tot}\lesssim 10$\% 
is applied, the effective temperature and luminosity, vary over very narrow intervals 
of values with varying metallicity, filling factor, and $\beta$ exponent as we may see 
from the first column and last row of frames in Fig. 2.
      
The effective temperatures \Teff\ from spectral type were calculated using the 
calibrations of \cite{Martins05} for O-type stars. A comparison of the numbers in columns 
3, 4, and 5 of Table 4 show that, in general, there is good agreement between the effective 
temperatures derived from FIT\textit{spec} and the values calculated from spectral type and other 
calibrations. The top plot of Fig. 4 compares the effective temperatures obtained from
FIT\textit{spec} (triangles) with the spectral type calibrations (asterisks) and the analyses
of \cite{Martins15} (plus signs) and \cite{Simon17} (squares) for our sample of stars. The
error bars measure the uncertainties in \Teff\ for each star in these calibrations. The
largest error bar corresponds to a temperature interval of 1.9 kK, while the shortest one 
corresponds to a length of 1 kK. The mean absolute errors between the \Teff\ data derived 
from FIT\textit{spec} and the corresponding data from spectral type and from all other 
calibrations in Table 4 are $\approx 1599$ K and $\approx 2338$ K, respectively. If we
compare the FIT\textit{spec} data to the more recent calibrations of \cite{Nieva13},
\cite{Martins15}, and \cite{Simon17} the mean absolute error decreases to $\approx 1946$ K.
When the uncertainties associated to the FIT\textit{spec} data are neglected, the sample 
standard deviation is $\approx 1331$ K, which is comparable to the uncertainty of 1000 K
in the FIT\textit{spec} data and the mean absolute deviations from the SpT - \Teff\
calibrations.

A further important validation of the FIT\textit{spec} code is the comparison of
the predicted surface gravities with the literature values. Columns 6, 7, and 8 of
Table 4 provide such a comparison with the SpT - \logg\ calibrations of \cite{Martins05}
and the results from the spectral analyses of \cite{Villamariz02}, \cite{Nieva13},
\cite{Martins15}, and \cite{Simon17}. There is also a general good agreement between the
FIT\textit{spec} data and the SpT - \logg\ calibrations. When the uncertainties in the
FIT\textit{spec} gravities are neglected, the mean absolute error between both sets of
values is $\approx 0.18$ dex. Similarly, the absolute deviation between the FIT\textit{spec}
gravities and the values listed in column 8 is $\approx 0.14$ dex, showing also a good
agreement with other calibrations in the literature. The scatter in the FIT\textit{spec}
data leads to a sample standard deviation of $\approx 0.20$ dex, which is {\bf above}
the uncertainty of $\pm 0.12$ dex in the predicted gravities {\bf and comparable to the
mean absolute error between the FIT\textit{spec} and SpT - \logg\ data}. The middle plot 
of Fig. 4 shows the comparison of the \logg\ values. The error bars depict the uncertainty 
($\pm 0.15$ dex) in the FIT\textit{spec} values and the calibrations of \cite{Martins15}.

The luminosities $L$ from spectral type are also calculated using the calibrations of
\cite{Martins05}. Columns 9 and 10 of Table 4 compare the luminosities derived from these
SpT - $L$ calibrations with those found by FIT\textit{spec}. We may see that the values 
calculated by FIT\textit{spec} are in very good agreement with those from the spectral
type calibrations, with absolute deviations varying from 0.02 to 0.37 dex. This
comparison is also displayed in the bottom plot of Fig. 4. The uncertainty in the
data as represented by the error bars is 0.15 dex for all stars and both calibrations.
The mean absolute error between both $log (L/L_{\odot})$ data sets is $\approx 0.16$
dex, which is very close to the actual uncertainty in the data. The largest deviation
from the spectral type luminosities occurs for star HD53975, with an absolute difference
of 0.37 dex. In addition, the luminosities calculated by FIT\textit{spec} exhibit a
{\bf dispersion with a sample standard deviation of $\approx 0.25$ dex, which is
almost twice the uncertainty in the FIT\textit{spec} luminosities}. 

It is well-known that the rotational broadening of unblended spectral lines changes the line
shape but does not affect the EW of the line \citep{Gray92}. Therefore, FIT\textit{spec}
does not need to apply rotational broadening before the adjustment of the effective 
temperature and gravity. In fact, this opens the possibility to estimate $v\cdot\sin i$ 
only with the rotational broadening by adjusting the synthetic spectra to the best fit of 
the observations, independently of \Teff\ and $\log g$. The last two columns of Table 4 
compare the results derived by such adjustments with those from several earlier and
more recent analyses. The mean absolute errors between both sets of data is 
$\approx 19.7$ km s$^{-1}$. This reasonable agreement demonstrates the reliability of the 
results for $v\cdot\sin i$. We may see from Table 4 that if the comparison is made
with the more recent calibrations of \cite{Oliveira06}, \cite{Nieva13}, \cite{Simon14},
\cite{Martins15}, and \cite{Simon17}, the mean absolute error {\bf increases} to $\approx 23.5$ 
km s$^{-1}$. {\bf This occurs mainly because of the rather large differences between
the FIT\textit{spec} data and the calibrations of \cite{Simon14} and \cite{Simon17}
for stars HD37022 and HD214680}.
In this work we have not taken into account the contribution of macroturbulence 
 or any other additional broadening mechanism in the determination of the projected 
rotational velocities. Any additional broadening mechanism will lower the contribution 
of rotational broadening for a given observation. In particular, \cite{Simon17} used 
high-resolution spectra of more than 400 stars
with spectral types in the range O4-B9 to provide new empirical clues to explain the
occurrence of macroturbulent spectral line broadening in O- and B-type massive stars.
They advanced the hypothesis that macroturbulent broadening may be the result of the
combined effects of pulsation modes associated with a heat-driven mechanism and 
possibly-cyclic motions originated by turbulent pressure instabilities, and concluded
that the latter mechanism could be the main responsible of the non-rotational
line broadening detected in OB stars. While the mechanisms proposed by \cite{Simon17}
still lack a definite confirmation, we may argue, based on the comparison between the
results of FIT\textit{spec} and the data of \cite{Simon17} for some of the stars in 
Table 4, that the effects of macroturbulent broadening are in fact those of lowering the 
projected rotational velocities. Finally, Figs. 5 to 9 compare the observed spectrum 
(black lines) with that derived from the best fit model (blue lines) for each star of 
our sample. Most of the salient spectral features are well reproduced by the models, 
showing the good quality of the fitting obtained by FIT\textit{spec}. Although
the results generated by FIT\textit{spec} are reliable, they can be improved by the
expert astronomer. In particular, they can be used in analyses where the parameters
of a large number of stars are required to be known or as the starting point to make
a better adjustment, especially in calibrations related to the chemical composition 
of the star. In any case, the use of FIT\textit{spec} represents a considerable
saving of time compared to other available tools.

The luminosity of a star is directly related to its mass and gravity, which are
directly reflected in the Balmer lines. The depth of these lines is in turn reflected
into their equivalent width, which is the main criterion employed by FIT\textit{spec}.
The use of this criterion has been demonstrated by the goodness of the fit when
comparing the effective temperatures and luminosities with those obtained from
SpT - $T_{\rm eff}$ and SpT - $L$ calibrations, respectively. 
In addition, the synthetic spectra of the 6D grid and the
FIT\textit{spec} code can be used to adjust observed spectra from a wide variety of
telescopes and spectrographs with different resolutions. A method of common use to
obtain the best automatic adjustment is to employ a chi-square ($\chi ^{2}$) statistics.
However, the appropriate use of a $\chi ^{2}$ test will degrade the synthetic spectra
at the resolution of the observation. Considering that the library of synthetic 
spectra currently consists of 2\,835\,000 spectra and that it will certainly continue 
to grow in number, a suitable comparison using the $\chi ^{2}$ statistics involves 
degrading the synthetic spectra at the same resolution of the observed spectrum. This
will also imply the use of additional CPU time. Although this is not a serious problem, 
it is completely avoided by using the comparison between the EWs and their ratios as 
the analysis technique. As a final remark, FIT\textit{spec} will be soon available for 
free download.

\section{Conclusions}

We have developed and tested the FIT\textit{spec} code, which uses a set of modern
automatic tools for searching the best fit models in a database consisting of 27\,000
{\sc cmfgen} model atmospheres. This database will be soon expanded to 80\,000 models.
 The code performs a quantitative spectroscopic analysis of large
samples of O- and B-type stars, using objective criteria in a fast and reliable
manner compared to traditional calibration tools. It effectively reduces the time
needed for the spectral analysis of massive OB stars from months to hours by 
identifying those models whose $Error_{\rm tot}$ is lower than the allowed tolerance
of $\lesssim 10$\% and discarding all those models that do not meet this criterion 
in order to find the effective temperature (\Teff) and the surface gravity ($\log g$) of 
a star by fitting the equivalent widths of optical He and \ion{H}{I} Balmer lines.

The reliability of the algorithm was assessed by analyzing the spectra of eight O-type
stars taken from the {\sc iacob} spectroscopic database of Northern Galactic OB stars
and comparing the derived results with those from spectral type - effective
temperature (SpT - \Teff), spectral type - surface gravity (SpT - \logg), and 
spectral type - luminosity (SpT - $L$) calibrations and from previous spectral analysis 
performed by other authors for the same stars.
The values of \Teff\ derived from FIT\textit{spec} are found to match well those 
calculated from SpT- \Teff\ calibrations and previous analyses from other authors,
with mean absolute errors of $\approx 1599$ K and $\approx 2338$ K, respectively.
The sample standard deviation of the data generated by FIT\textit{spec} is
$\approx 1331$ K, which is well within the range of the mean absolute deviations
from the SpT- \Teff\ and other calibrations in the literature. On the other hand,
 the values of the surface gravity derived by FIT\textit{spec} agree reasonably
well with those obtained from SpT -\logg\ calibrations, with a mean absolute error
of $\approx 0.18$ dex. A {\bf lower} absolute deviation of $\approx 0.14$ dex was obtained
by comparing with other calibrations. The values of the stellar luminosity derived by 
the FIT\textit{spec} algorithm were also found to agree with those obtained from 
the SpT - $L$ calibrations, with a mean absolute error of $\approx 0.16$ dex. This 
deviation from the SpT - $L$ calibrations is comparable to the uncertainty of 0.15 dex 
in the FIT\textit{spec} data, which appears to be independent of the spectral type at 
least for the stars considered in this study.

In order to complement the database of stellar atmosphere models, we have also developed
a library of rotationally broadened synthetic spectra, which allows quick estimation 
of the projected rotational velocity ($v\cdot\sin i$) of a star. The results of the
adjustments using this library are also found to agree reasonably well with results
from other spectroscopic analyses for the same stars, with a mean absolute error
of $\approx 19.7$ km s$^{-1}$ when earlier and recent calibrations are taken into
account. If the data is compared only with the more recent calibrations, the mean
absolute error {\bf increases} to $\approx 23.5$ km s$^{-1}$. The good agreement of the
results obtained from FIT\textit{spec} with other spectral analyses demonstrates the
reliability of the models.

\section{Acknowledgments}

We thank the referee for providing a number of valuable comments and 
suggestions that have improved the content of the manuscript.
We acknowledge support from the {\sc abacus}-Centro de Matem\'atica Aplicada y 
C\'omputo de Alto Rendimiento of Cinvestav-IPN under grant EDOMEX-2011-C01-165873.
One of us J.Z. is grateful for support by the CONACyT project CB-2011-01 No. 168632.
The calculations of this paper were performed using the {\sc abacus} computing
facilities.

\clearpage


\begin{deluxetable}{cl}
\tablecaption{Stellar parameters \label{tab:parameters}}
\tablecolumns{2}
\tablewidth{0pt}
\tablehead{
\colhead{Parameters in 6D space} &
\colhead{Value}
}
\startdata
\Teff     & from evolutive tracks\tablenotemark{a} \\
$L$       & from evolutive tracks\tablenotemark{a} \\
$Z$       & solar metallicity and solar metallicity enhanced \\
          & by rotation from evolutive tracks\tablenotemark{a} \\
$v_\infty$ & 2.1$v_{\rm esc}$\\
$\beta$   & 0.5, 0.8, 1.1, 1.4, 1.7, 2.1, 2.3\\
$F_{\rm cl}$   & 0.05, 0.30, 0.60, 1.00\\
\hline
Other Parameters & Value\\
\hline
$M$       & from evolutive tracks\tablenotemark{a} \\
$R$       & from $M$ and $L$\\
\logg     & from $M$ and $L$\\
\vsini    & from library of synthetic spectra \\
$\dot{M}$ & from evolutive tracks\tablenotemark{a} \\
\enddata
\tablenotetext{a}{\cite{Ekstrom12}.}
\end{deluxetable}

\clearpage


\begin{deluxetable}{ccccccccc}
\tablecaption{Super levels/levels for the different ionization stages included in the models.
\label{tab:superlevels}}
\tablehead{
\colhead{Element} & \colhead{I} &
\colhead{II} & \colhead{III} & \colhead{IV} &
\colhead{V} & \colhead{VI} & \colhead{VII} &
\colhead{VIII}
}
\startdata
H~  & 20/30   & 1/1~~~  & \nodata  & \nodata & \nodata & \nodata & \nodata & \nodata\\
He  & 45/69   & 22/30~  & 1/1~~~~~ & \nodata & \nodata & \nodata & \nodata & \nodata\\
C~  & \nodata & 40/92~  & 51/84~~~ & 59/64~  & 1/1~~~  & \nodata & \nodata & \nodata\\
N~  & \nodata & 45/85~  & 41/82~~~ & 44/76~  & 41/49~  & 1/1~~~  & \nodata & \nodata\\
O~  & \nodata & 54/123  & 88/170~~ & 38/78~  & 32/56~  & 25/31~  & 1/1~~~  & \nodata\\
Si  & \nodata & \nodata & 33/33~~~ & 22/33~  & 1/1~~~  & \nodata & \nodata & \nodata\\
P~  & \nodata & \nodata & \nodata  & 30/90~  & 16/62~  & 1/1~~~  & \nodata & \nodata\\
S~  & \nodata & \nodata & 24/44~~~ & 51/142  & 31/98~  & 28/58~  & 1/1~~~  & \nodata\\
Fe  & \nodata & \nodata & 104/1433 & 74/540  & 50/220  & 44/433  & 29/153  &  1/1 \\
\enddata
\end{deluxetable}

\clearpage


\begin{deluxetable}{cccccccccccc}
\tablecaption{Parameters of the best fit models found by FIT\textit{spec}.
\label{tab:parameters}}
\tablehead{
\colhead{Star} & \colhead{\Teff} & \colhead{$log(\frac{L}{L_{\odot}})$} & \colhead{$M$} & \colhead{$R$} &
\colhead{\logg} & \colhead{$Z$} & \colhead{$\dot{M}$} & 
 \colhead{$v_{\infty}$} &
\colhead{$F_{\rm cl}$} & \colhead{$\beta$} &
\colhead{$v\cdot\sin i$}\\
\colhead{} & \colhead{(K)} & \colhead{} & \colhead{($M_{\odot}$)} &
	\colhead{($R_{\odot}$)} & \colhead{(cm s$^{-2}$)} & \colhead{} & 
\colhead{($M_{\odot}$ yr$^{-1}$)} & \colhead{(km s$^{-1}$)} &
\colhead{} & \colhead{} & \colhead{(km s$^{-1}$)}
}
\tabletypesize{\tiny}
\startdata
HD34078~ & 33\,580$\pm$1000 & 4.66$\pm$0.15 & 18.99 & 43.25 & 4.120$\pm$0.12 &Sun\tablenotemark{a} &7.055$\times 10^{-9}$&2\,260 & 0.05 & 1.1 & ~30 \\
HD36512~ & 31\,280$\pm$1000 & 4.40$\pm$0.15 & 15.70 & 37.25 & 4.168$\pm$0.12 &Sun\tablenotemark{a} &7.808$\times 10^{-9}$&2\,220 & 0.30 & 1.7 & ~30 \\
HD36879~ & 32\,220$\pm$1000 & 5.25$\pm$0.15 & 25.04 & 91.90 & 3.572$\pm$0.12 &SER\tablenotemark{b} &2.063$\times 10^{-7}$&1\,770 & 0.30 & 0.8 & 180 \\
HD37022~ & 33\,470$\pm$1000 & 4.91$\pm$0.15 & 21.66 & 58.37 & 3.913$\pm$0.12 &SER\tablenotemark{b} &7.262$\times 10^{-8}$&2\,070 & 0.60 & 0.5 & 100 \\
HD53975~ & 35\,030$\pm$1000 & 4.63$\pm$0.15 & 19.86 & 38.76 & 4.236$\pm$0.12 &Sun\tablenotemark{a} &6.005$\times 10^{-9}$&2\,440 & 0.05 & 1.4 & 160 \\
HD54662~ & 35\,500$\pm$1000 & 4.90$\pm$0.15 & 22.75 & 50.65 & 4.060$\pm$0.12 &Sun\tablenotemark{a} &1.890$\times 10^{-8}$&2\,280 & 0.05 & 0.5 & ~80 \\
HD193322 & 32\,460$\pm$1000 & 4.74$\pm$0.15 & 19.05 & 50.67 & 3.982$\pm$0.12 &Sun\tablenotemark{a} &2.336$\times 10^{-8}$&2\,090 & 0.30 & 1.7 & ~50 \\
HD214680 & 32\,980$\pm$1000 & 4.66$\pm$0.15 & 18.62 & 44.97 & 4.077$\pm$0.12 &Sun\tablenotemark{a} &1.793$\times 10^{-8}$&2\,930 & 0.30 & 1.1 & ~40 \\
\enddata
\tablenotetext{a}{Solar metallicity: H, He, C, N, and, O from evolutive tracks of \cite{Ekstrom12}
and Si, P, S and Fe from \cite{Asplund09}}
\tablenotetext{b}{Solar metallicity enhanced by rotation: H, He, C, N, and, O from evolutive 
tracks of \cite{Ekstrom12}and Si, P, S and, Fe from \cite{Asplund09}}
\end{deluxetable}

\clearpage


\rotate
\begin{deluxetable}{ccccccccccccc}
\tablecaption{Comparison of parameters found by FIT\textit{spec}
with the results of previous calibrations.
\label{tab:compare_parameters}}
\tablehead{
\colhead{Star} & \colhead{SpT} & \colhead{\Teff} & \colhead{\Teff} &
\colhead{\Teff} & \colhead{$\logg$} & \colhead{$\logg$} & \colhead{$\logg$} &
\colhead{$\log (\frac{L}{L_{\odot}})$} & \colhead{$\log (\frac{L}{L_{\odot}})$} &
\colhead{$v\cdot\sin i$} & \colhead{$v\cdot\sin i$}\\
\colhead{} & \colhead{} & \colhead{(K)} & \colhead{(K)} & \colhead{(K)} &
\colhead{(cm s$^{-2}$)} & \colhead{(cm s$^{-2}$)} & \colhead{(cm s$^{-2}$)} &
\colhead{} & \colhead{} & \colhead{(km s$^{-1}$)} & \colhead{(km s$^{-1}$)} \\
\colhead{} & \colhead{} & \colhead{FIT{\it spec}} & \colhead{SpT} & \colhead{other} &
\colhead{FIT{\it spec}} & \colhead{SpT} & \colhead{other} &
\colhead{FIT{\it spec}} & \colhead{SpT} & \colhead{FIT{\it spec}} &
\colhead{other}
}
\tabletypesize{\tiny}
\startdata
HD34078~ & O9.5V & 33\,580$\pm$1000 & 30\,488$\pm$1000 & 33\,000$^{a}$~~~~~~ & 4.120$\pm$0.12 & 3.92 & 4.0$\pm$0.15$^{a}$ & 4.66$\pm$0.15 & 4.62$\pm$0.15 & ~30 & ~25$^{a}$ \\
	 &       &         &                    & ~~33\,900$\pm$1700$^{b}$ & & & 3.980$^{b}$~~~~~~&  &  &  & ~17$^{b}$ \\
	 &       &         &                    & ~~36\,500$\pm$1000$^{c}$ & & &~4.05$^{c}$~~~~~~~~~& &  &  & ~40$^{c}$ \\
	 &       &         &                    &                          & & &                   & &  &  & ~27$^{f}$ \\
	 &       &         &                    &                          & & &                   & &  &  & ~13$^{i}$ \\
\hline
HD36512~ & O9.7V & 31\,280$\pm$1000 & 30\,000$^{d}$~~~~&32\,500$^{a}$~~~~~~~ & 4.168$\pm$0.12 & 3.92$^{d}$ & 4.0$\pm$0.15$^{a}$ & 4.40$\pm$0.15 & 4.58$^{d}$& ~30 & ~20$^{a}$ \\
         &       &         &                    & ~33\,900$\pm$1700$^{b}$ & & &~4.210$^{b}$~~~~~~ & &  &  & ~15$^{b}$ \\
         &       &         &                    & ~33\,400$\pm$~200$^{e}$  & & &~~4.30$\pm$0.05$^{e}$& &      &    & ~20$\pm$2$^{e}$ \\
         &       &         &                    &                          & & &4.13~$^{f}$~~~~~ & &      &    & ~15$^{i}$ \\
\hline
HD36879~ & ~~O7V & 32\,220$\pm$1000 & 35\,531$\pm$1000 & 36\,500$^{a}$~~~~~~ &3.572$\pm$0.12 & 3.92 & 3.75$\pm$0.15$^{a}$ & 5.25$\pm$0.15 & 5.10$\pm$0.15 & 180 & 200$^{a}$ \\
	 &       &                    &                    &                     & & & &      &      &     & 200$^{f}$ \\
	 &       &                    &                    &                     & & & &      &      &     & 219$^{i}$ \\
\hline
HD37022~ & ~~O7V & 33\,470$\pm$1000 & 35\,531$\pm$1000 & ~~38\,900$\pm$1700$^{b}$ & 3.913$\pm$0.12 & 3.92 & 4.170$^{b}$~~~~~~& 4.91$\pm$0.15 & 5.10$\pm$0.15 & 100 & ~26$^{b}$ \\
	 &       &         &                    &                          & & &                   & &  &  & ~98$^{f}$ \\
	 &       &         &                    &                          & & &                   & &  &  & ~26$^{i}$ \\
\hline
HD53975~ & O7.5V & 35\,030$\pm$1000 & 34\,419$\pm$1000 & ~~35\,500$\pm$1900$^{b}$ & 4.236$\pm$0.12 & 3.92 & 3.590$^{b}$~~~~~~& 4.63$\pm$0.15 & 5.00$\pm$0.15 & 160 & 186$^{b}$ \\
	 &       &                    &                    & 36\,300$^{h}$~~~~~~     & & & &      &      &            &147$^{f}$ \\
         &       &                    &                    &                         & & & &      &      &            &163$^{g}$ & \\
	 &       &                    &                    &                         & & & &      &      &            &180$^{i}$ \\
\hline
HD54662~ & ~~O7V & 35\,500$\pm$1000 & 35\,531$\pm$1000 & &4.060$\pm$0.12 & 3.92 &         & 4.90$\pm$0.15 & 5.10$\pm$0.15 & ~80 & ~70$^{f}$ \\
\hline
HD193322 & ~~O9V & 32\,460$\pm$1000 & 31\,524$\pm$1000 & & 3.982$\pm$0.12 & 3.92 &          & 4.74$\pm$0.15 & 4.72$\pm$0.15 & ~50 & ~94$^{f}$ \\
	 &       &                    &                    & &&&         &               &               &     & ~41$^{i}$ \\
\hline
HD214680 & ~~O9V & 32\,980$\pm$1000 & 31\,524$\pm$1000 & 35\,000$^{a}$~~~~~~ & 4.077$\pm$0.12 & 3.92 & 4.05$\pm$0.15$^{a}$ & 4.66$\pm$0.15 & 4.72$\pm$0.15 & ~40 & ~15$^{a}$ \\
	 &       &                    &                    & ~~35\,500$\pm$1900$^{b}$ & & & 3.920$^{b}$~~~~~~~&      &  & &~16$^{b}$ \\
	 &       &                    &                    & ~~37\,500$\pm$1000$^{c}$ & & & 4.0$^{c}$~~~~~~~~~~&      &  & &~50$^{c}$ \\
	 &       &                    &                    &                          & & &                    &      &  & & ~32$^{f}$ \\
	 &       &                    &                    &                          & & &                    &      &  & & ~16$^{i}$ \\
\enddata
\tablenotetext{a}{\cite{Martins15}}
\tablenotetext{b}{\cite{Simon17}}
\tablenotetext{c}{\cite{Villamariz02}}
\tablenotetext{d}{Extrapolated using the data of Table 1 of \cite{Martins05}}
\tablenotetext{e}{\cite{Nieva13}}
\tablenotetext{f}{\cite{Conti77}}
\tablenotetext{g}{\cite{Oliveira06}}
\tablenotetext{h}{\cite{Sybesma82}}
\tablenotetext{i}{\cite{Simon14}}
\end{deluxetable}

\clearpage


\clearpage

\begin{figure}
\epsscale{0.8}
\plotone{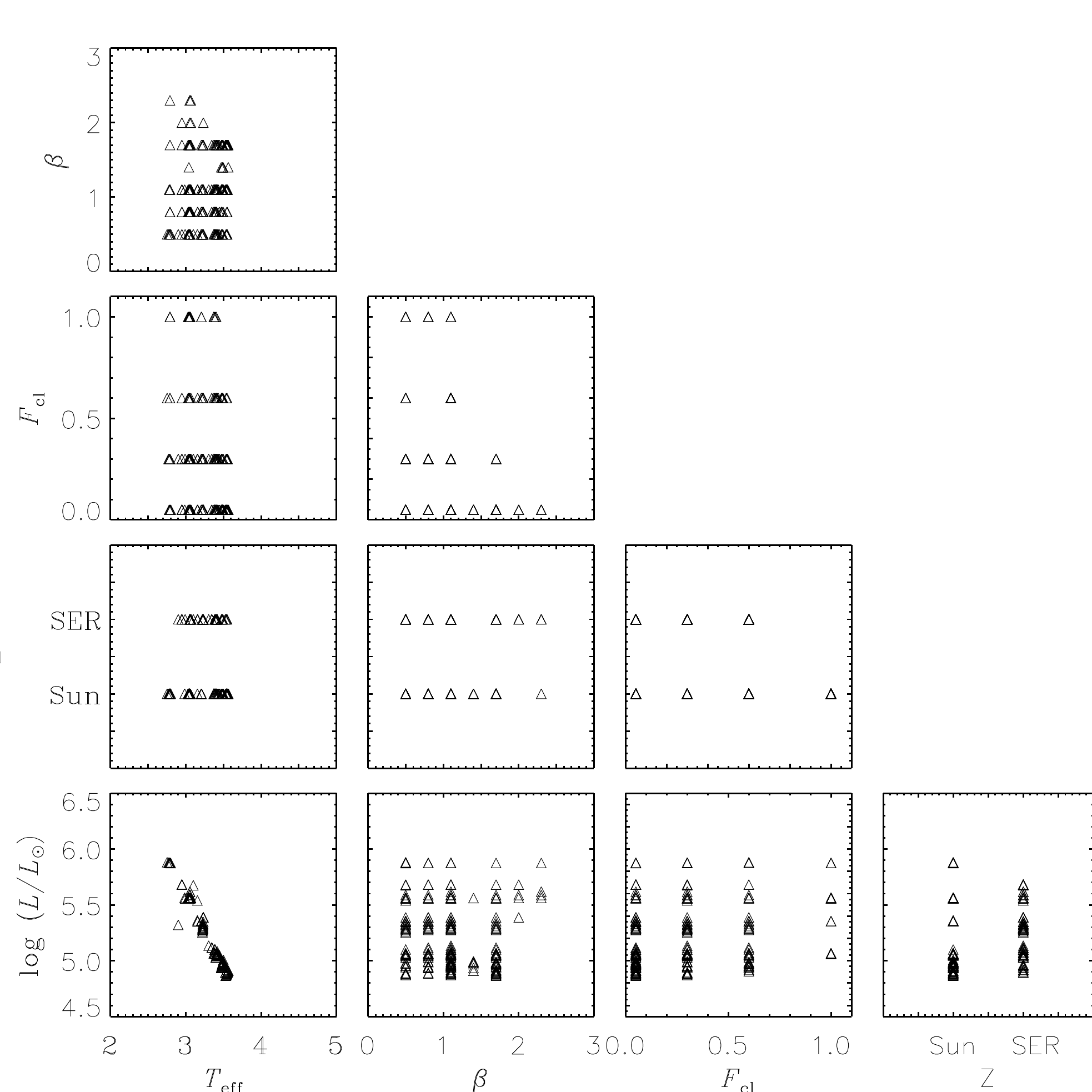}
\caption{Distribution of the models with errors less than 50\% in the six-dimensional (6D) parameter space 
for star HD54662 of spectral type O7 V. In the first frame of the third row the legend Sun refers to solar 
metallicity, while SER refers to solar metallicity enhanced by rotation. 
\label{Fig. 1}}
\end{figure}
\clearpage
\begin{figure}
\epsscale{0.8}
\plotone{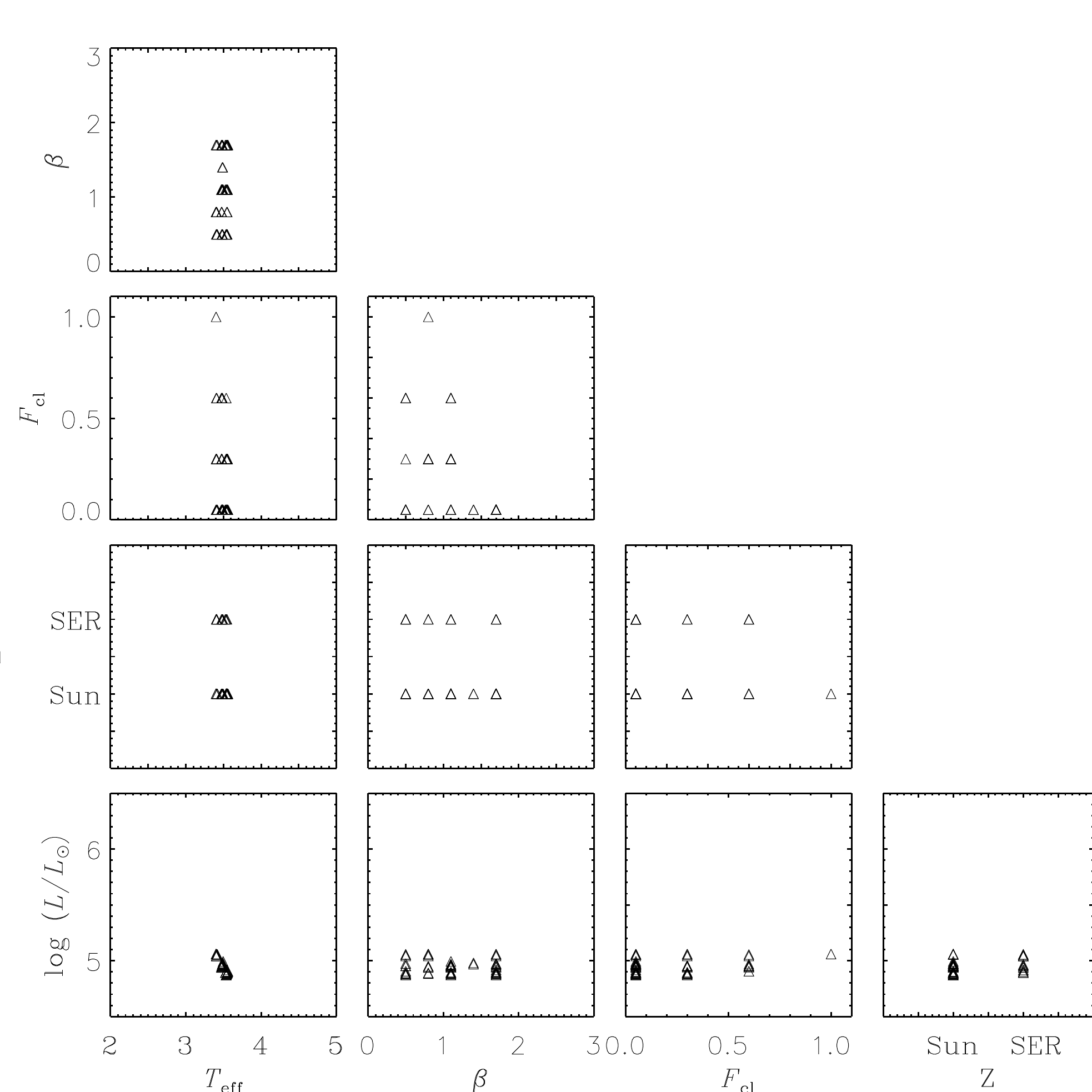}
\caption{Distribution of the best fit models (with errors less than 10\%) in the six-dimensional (6D) parameter 
space for star HD54662. In the first frame of the third row the legend Sun refers to solar metallicity, while 
SER refers to solar metallicity enhanced by rotation.
\label{Fig. 2}}
\end{figure}
\clearpage
\begin{figure}
\epsscale{0.8}
\plotone{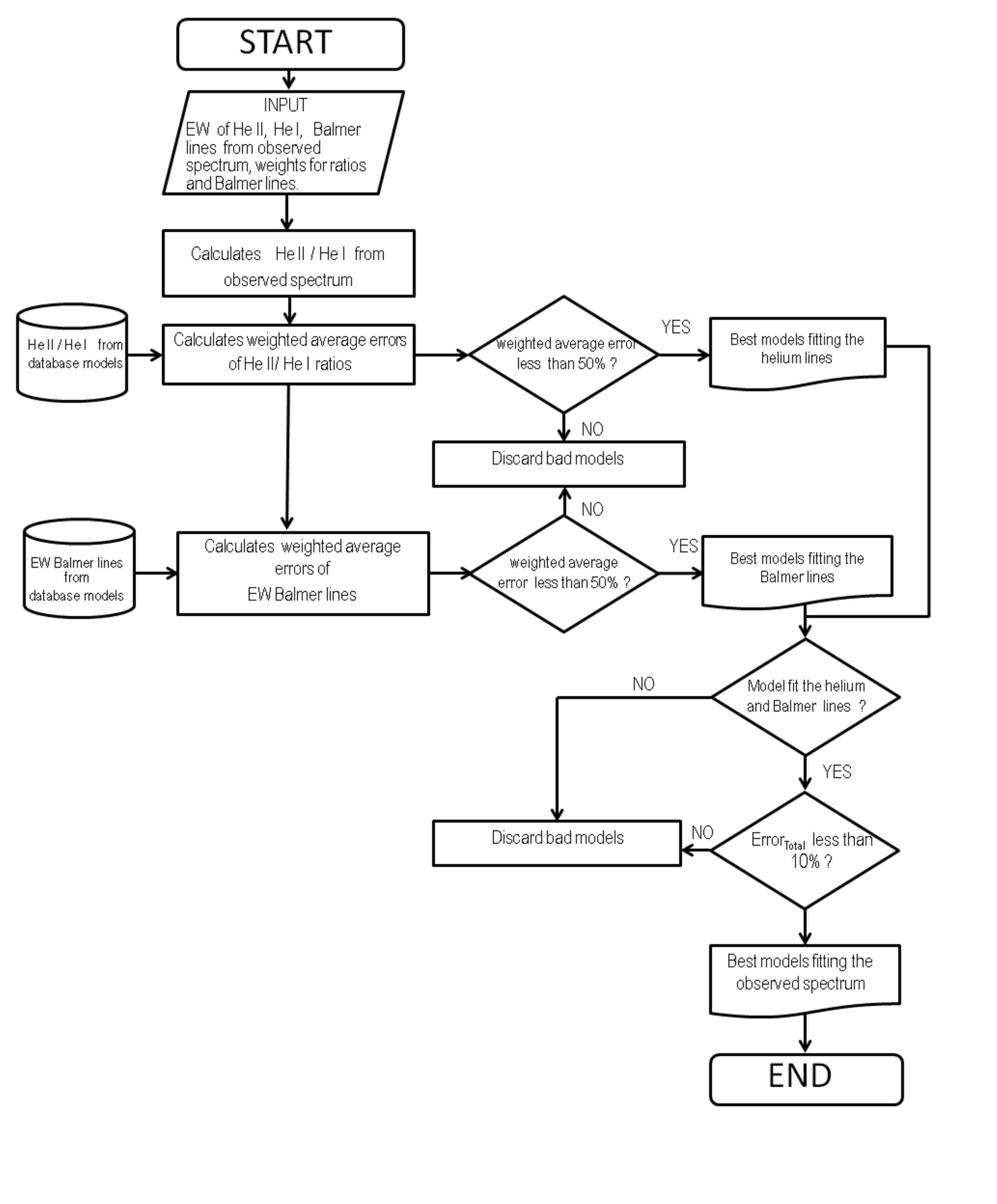}
\caption{Flowchart showing the sequence of steps followed by the FIT\textit{spec} algorithm.
\label{Fig. 3}}
\end{figure}
\clearpage
\begin{figure}
\epsscale{0.7}
\plotone{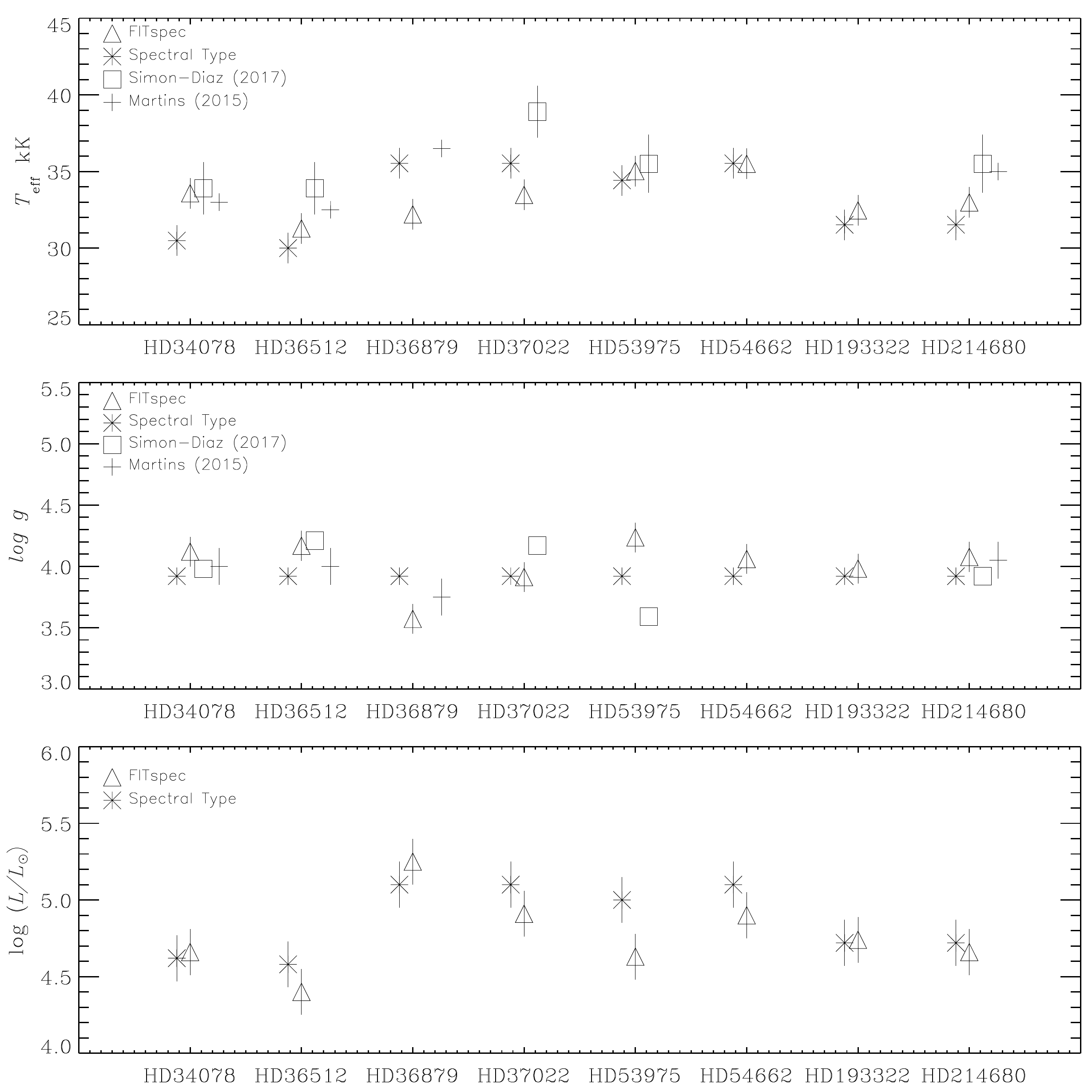}
\caption{(Top panel) Effective temperatures found by FIT\textit{spec} (triangles) compared 
to the corresponding values obtained from the spectral type (asterisks) and the adjustments of 
of \cite{Martins15} (plus signs) and \cite{Simon17} (squares). The largest error bar 
corresponds to a temperature interval of 1.9 kK, while the shortest one corresponds to a 
length of 1 kK. (Middle panel) Surface gravity found by FIT\textit{spec} (triangles)
compared to the corresponding values obtained from spectral type (asterisks) and the
adjustments of \cite{Martins15} (plus signs) and \cite{Simon17} (squares). The length of the
error bars are 0.12 dex for the FIT\textit{spec} data and 0.15 dex for the calibrations of
\cite{Martins15}. (Bottom panel) Logarithm of the luminosity found by FIT\textit{spec} 
(triangles) compared to the corresponding values obtained from the spectral type (asterisks). 
All error bars correspond to 0.15 dex.
\label{Fig. 4}}
\end{figure}
\clearpage
\begin{figure}
\epsscale{0.6}
\plotone{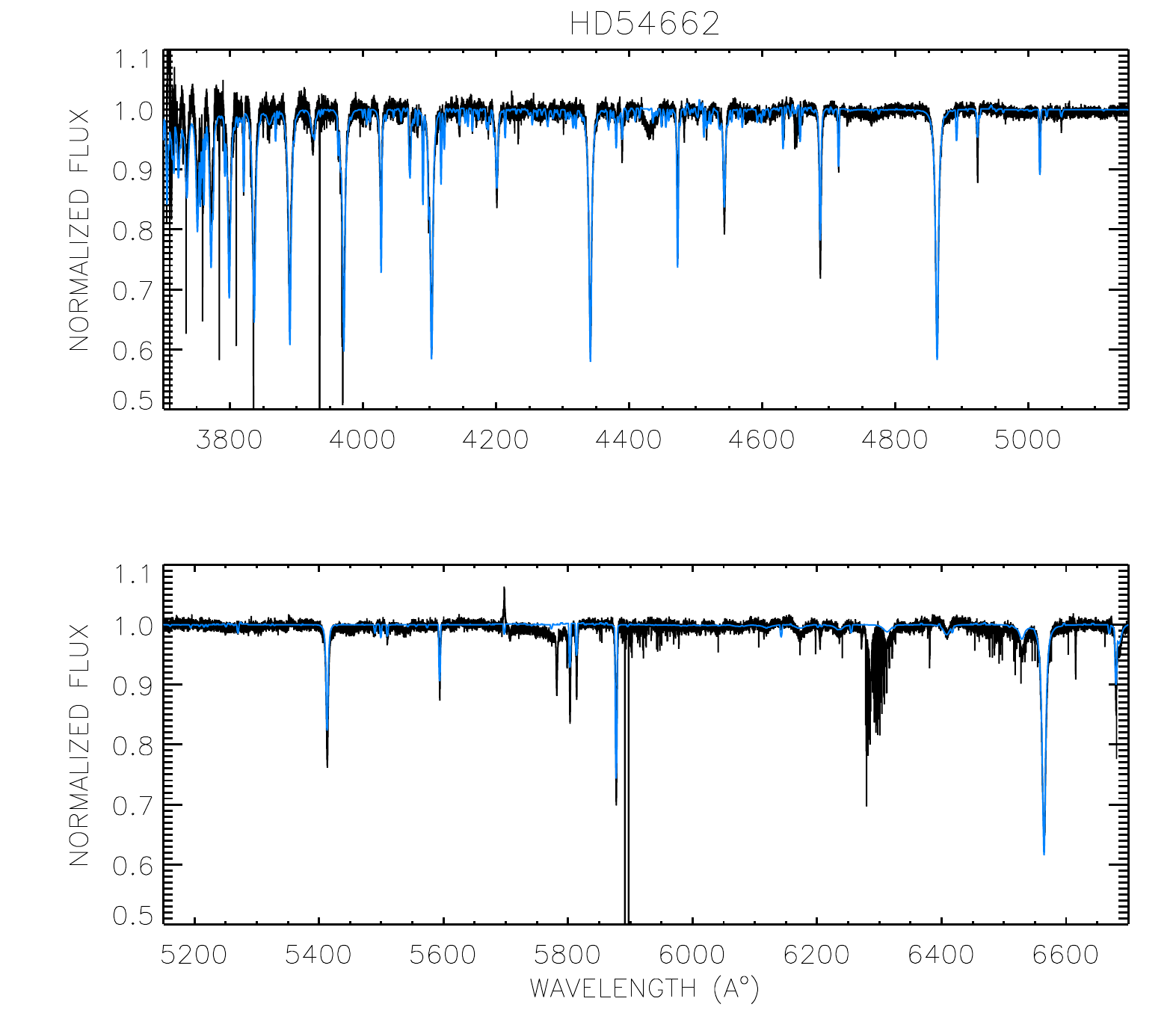}
\caption{Comparison of the observed spectrum (black line) with that of the
best fit model (blue line) for star HD54662.
\label{Fig. 5}}
\end{figure}
\clearpage
\begin{figure}
\epsscale{0.6}
\plotone{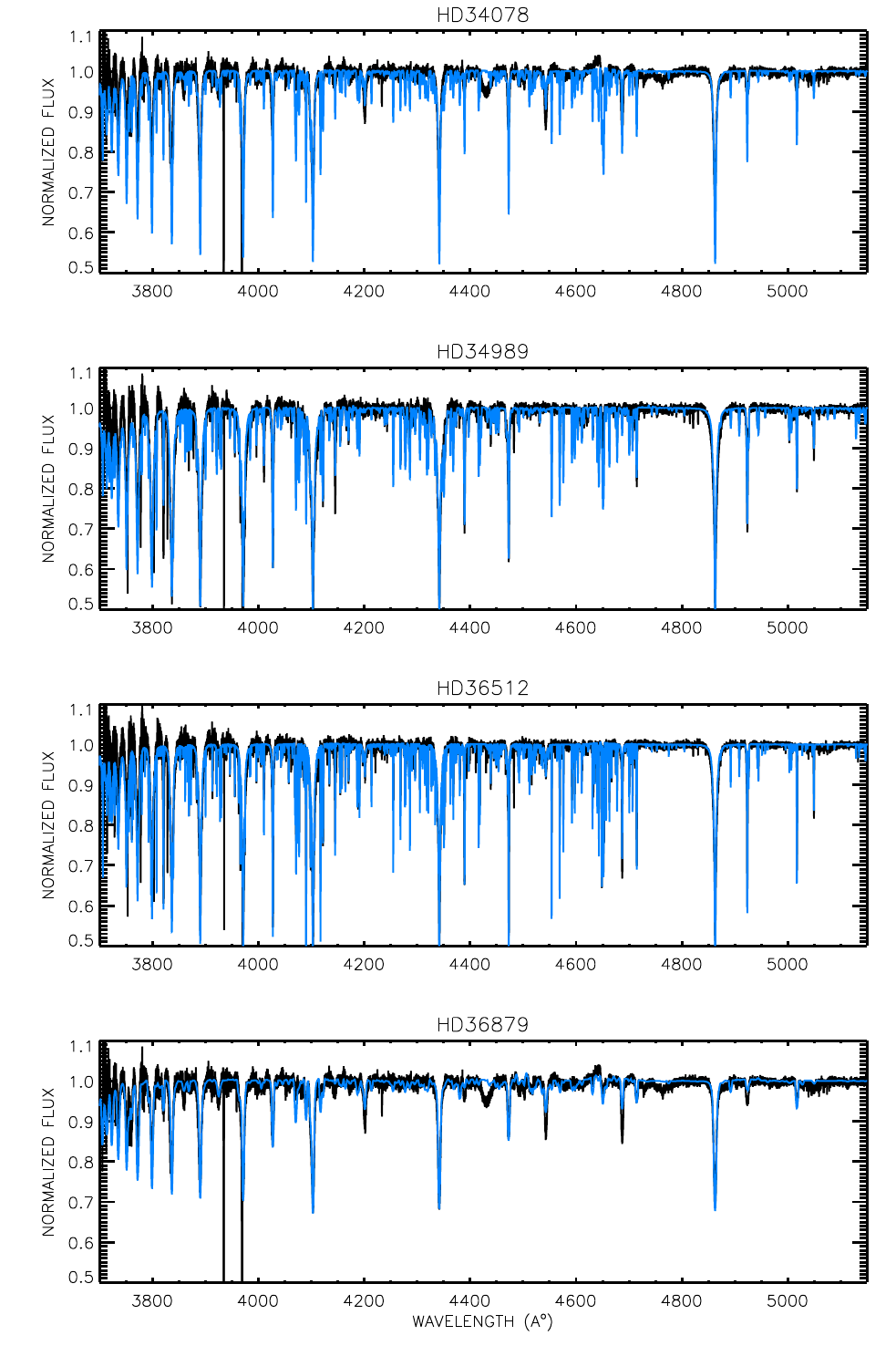}
\caption{Comparison of the observed spectrum (black line) with that of the
best fit model (blue line) for selected sample of stars. 
\label{Fig. 6}}
\end{figure}
\clearpage
\begin{figure}
\epsscale{0.6}
\plotone{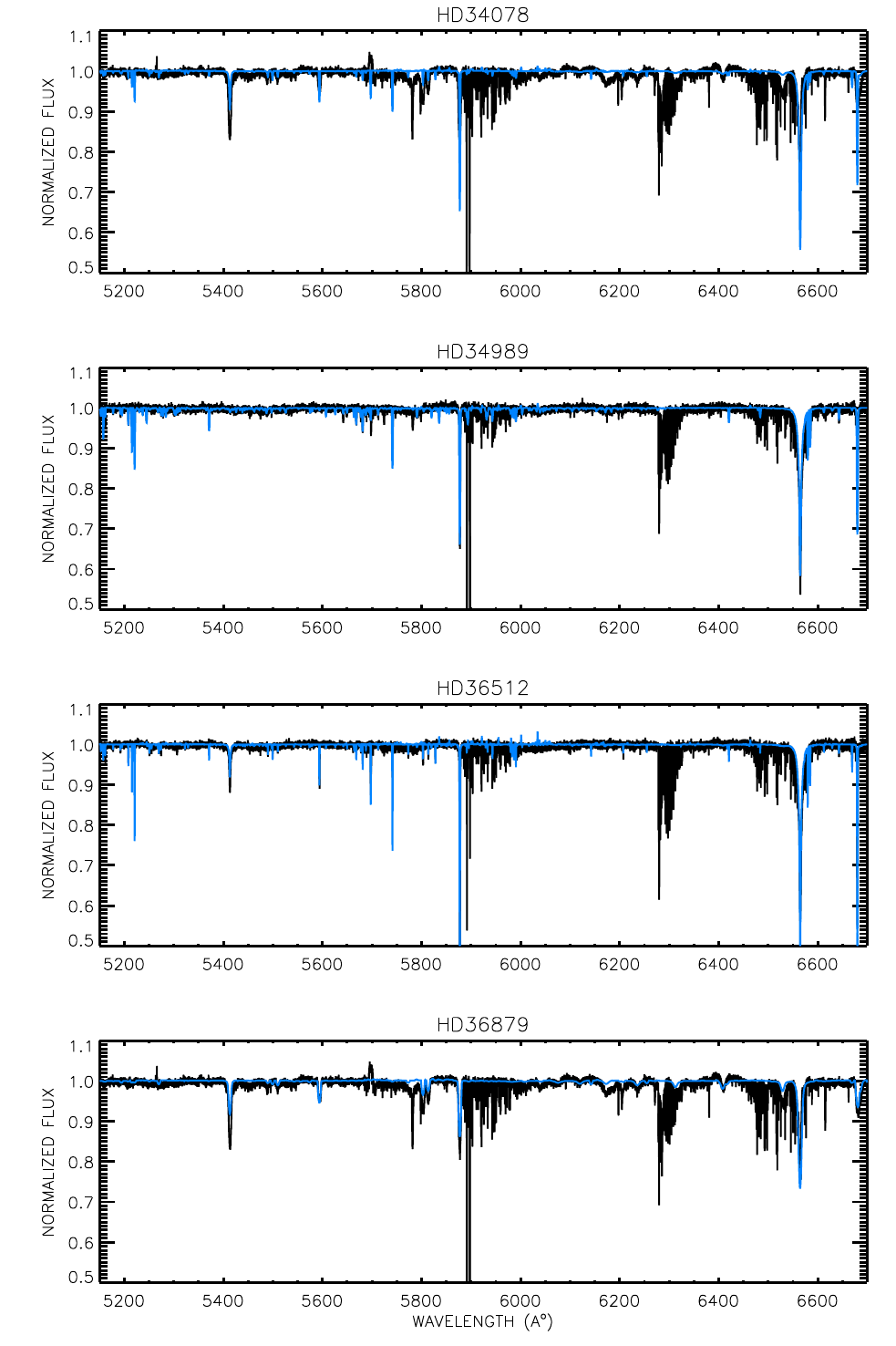}
\caption{Comparison of the observed spectrum (black line) with that of the
best fit model (blue line) for selected sample of stars. 
\label{Fig. 7}}
\end{figure}
\clearpage
\begin{figure}
\epsscale{0.6}
\plotone{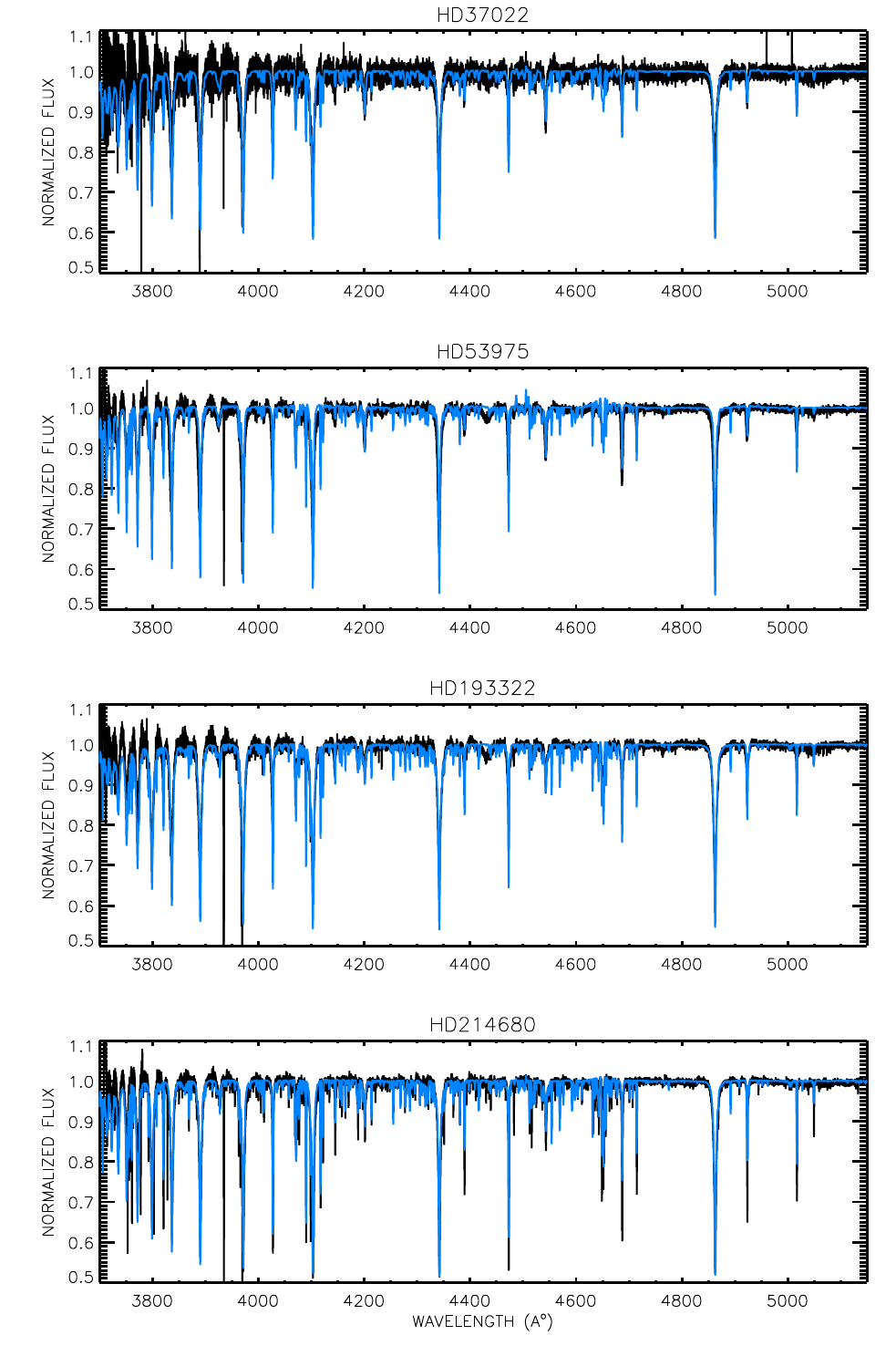}
\caption{Comparison of the observed spectrum (black line) with that of the
best fit model (blue line) for selected sample of stars. 
\label{Fig. 8}}
\end{figure}
\clearpage
\begin{figure}
\epsscale{0.6}
\plotone{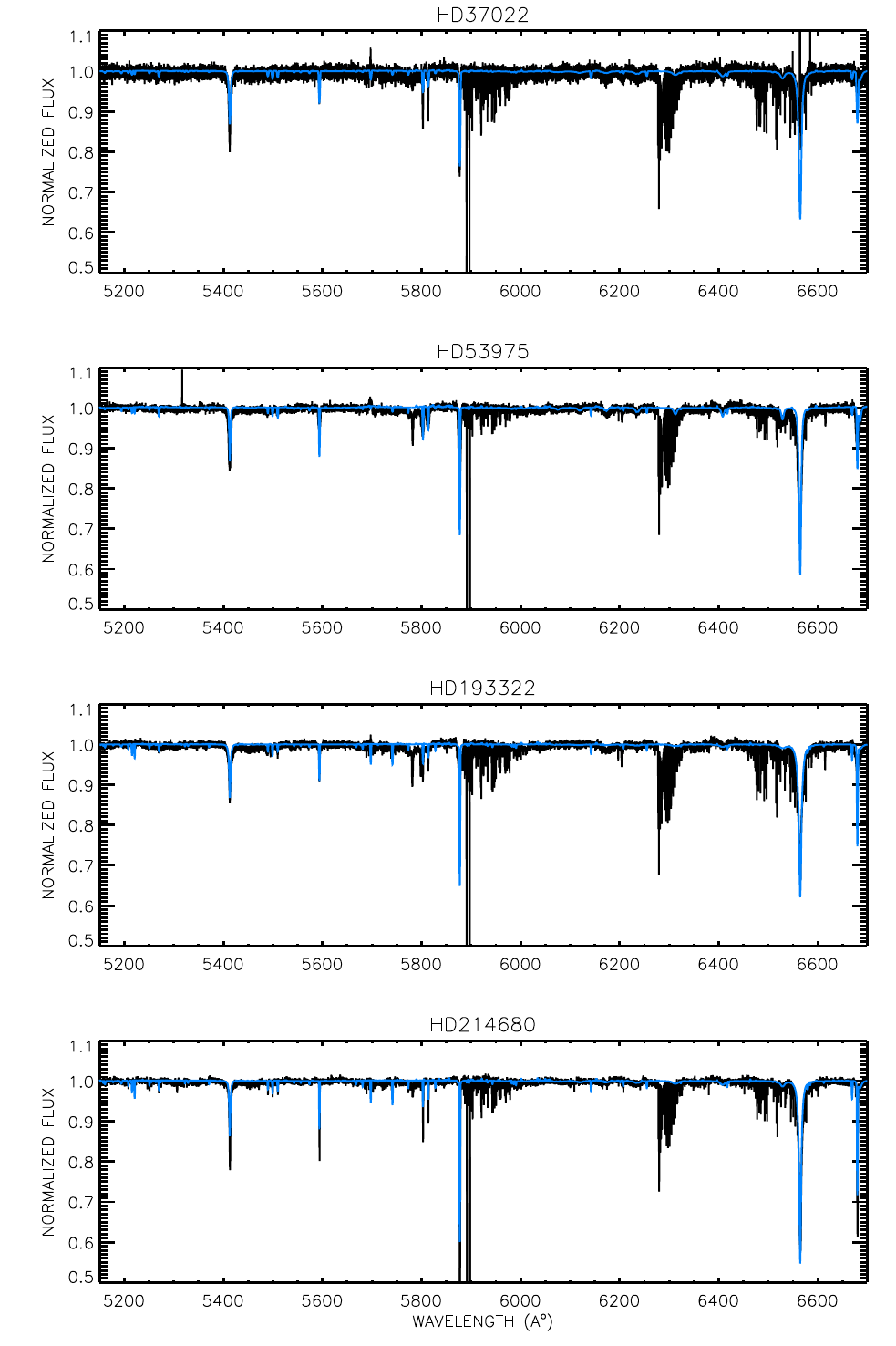}
\caption{Comparison of the observed spectrum (black line) with that of the
best fit model (blue line) for selected sample of stars. 
\label{Fig. 9}}
\end{figure}

\end{document}